\documentstyle[prl, twocolumn,aps]{revtex}
\begin{document}
\begin{title}
{Quasiparticle lifetimes in the charged Bose gas and the cuprates}
\end{title} 
\author{A.S. Alexandrov and C. J. Dent}
\address
{Department of Physics, Loughborough University, Loughborough LE11 
3TU, U.K.}

\maketitle
\begin{abstract}
The scattering cross section of a Coulomb potential
screened by a charged Bose gas (CBG) is calculated both above and below the
Bose-Einstein condensation temperature, using the variable phase
method. In contrast with the BCS superconductor, the screened
scattering potential
 and quasiparticle lifetime are found to be very different in
the superconducting and normal states. We apply the result to explain
 the appearence of a sharp peak in
the ARPES spectra in some cuprates below the superconducting transition.
\end{abstract}
\narrowtext

There is a growing body of evidence that cuprate superconductivity is
due to the condensation of bipolarons, local bosonic pairs of carriers
bound by the strong electron-phonon interaction \cite{alemot}.
The theory has been applied to explain the upper critical field \cite{osofsky},
magnetic susceptibility \cite{muller}, anisotropy \cite{hofer},
isotope effect on the supercarrier mass \cite{zhao} and the pseudogap
\cite{alemot,mihailovic,rubio}. It provides a parameter-free formula 
for the superconducting
$T_c$ \cite{alekab} and a parameter-free fit to the electronic
specific heat near the transition \cite{alebee}. The d-wave 
order parameter and the
 single particle tunneling density of
states can be understood in the framework of Bose-Einstein
condensation of inter-site bipolarons as well \cite{aletun,aleand}. 
We have also explained various features of the data from
angle-resolve photoemission spectroscopy (ARPES) \cite{aleden}. We assumed that 
a single photoexcited hole in the oxygen band is scattered by
impurities, while the chemical potential is pinned inside the 
charge-transfer (optical) gap due to bipolaron formation.
The normal state gap, the spectral shape and the polarisation
dependence of the ARPES spectra were well described within this approach in
a few cuprates.
 Recently it has been observed at certain points
in the Brillouin zone that the ARPES peak in the bismuth cuprates
is relatively sharp at low
temperatures in the superconduncting state, but that it almost disappears 
into the background above the
transition \cite{kam,fed}. 

In this letter, we first calculate the
scattering cross section of single-particle excitations off
a Coulomb scattering centre in the charged
Bose gas, both above and below the Bose-Einstein condensation temperature.
We then
propose that the appearence of a sharp ARPES peak below the transition 
in the cuprates is caused by a large increase in the quasiparticle lifetime 
due to condensate screening of scatterers.

First we calculate the scattering cross-section of a charged
particle (mass $m$, charge $e$) scattered by a static Coulomb potential
$V(r)$ screened by the CBG. 
The general theory of potential scattering in terms of phase shifts
was developed in the earliest days of quantum mechanics (see for
instance \cite{lanlif}.) While in principle this allows scattering
cross sections to be calculated for an arbitary potential, in practice
the equations for the radial part of the wavefunction may only be
solved analytically for a few potentials, and in the standard
formulation are not in a suitable form for numerical computation. The
`variable phase' approach \cite{calogero} solves this problem by
making the phase shifts functions of the radial coordinate, and then
the Schrodinger equation for each radial component of the wavefunction
reduces to a first order differential equation for the corresponding
phase shift.

In dimensionless units ($\hbar=2m=1$), the Schrodinger equation for
 the radial part of 
 the angular momentum $l$ component of the wavefunction of a 
particle with wavevector $k$ undergoing potential scattering is
     \begin{equation}
          u_{l}''(r)+\left[k^{2}-l(l+1)/r^{2}-V(r)\right]u_{l}(r)=0.
     \end{equation}
The scattering phase shift $\delta_{l}$ is obtained by comparison 
with the asymptotic relation
     \begin{equation}
          u_{l}(r)\stackrel{r\rightarrow\infty}{\longrightarrow}
               \sin(kr-l\pi/2+\delta_{l}),
     \end{equation}
and the scattering cross section is then
     \begin{equation}
          \sigma={4\pi\over{k^{2}}}\sum_{l=0}^{\infty}\sin^{2}\delta_{l}.
     \end{equation}
In the variable phase method \cite{calogero}, we must satisfy the condition 
 that
     \begin{equation}
          V(r)\stackrel{r\rightarrow0}{\longrightarrow}V_{0}r^{-n},
     \end{equation}
with $n<2$. The angular momentum $l$ phase shift is then
     \begin{equation}
          \delta_{l}=\lim_{r\rightarrow\infty}\delta_{l}(r)
     \end{equation}
where the phase function $\delta_{l}(r)$ satisfies the phase equation
     \begin{equation}
          \delta_{l}'(r)=-k^{-1}V(r)\left
[\cos\delta_{l}(r)\hat{j}_{l}(kr)-\sin\delta_{l}(r)\hat{n}_{l}(kr)\right]^{2},
     \end{equation}
with
     \begin{equation}
          \delta_{l}(r)\stackrel{r\rightarrow0}{\longrightarrow}
-{V_{0}r^{-n}\over{k^{2}}} {(kr)^{2l+3}\over{(2l+3-n)[(2l+1)!!]^{2}}},
     \end{equation}
and $j_l(x)$ and $n_l(x)$ are the Riccati-Bessel functions \cite{calogero}.
In the $l=0$ case, 
the phase equation reduces to
     \begin{equation}
          \delta_{0}'(r)=-k^{-1}V(r)\sin^{2}[kr+\delta_{0}(r)].
     \end{equation}
In the slow particle limit, we may also
neglect higher order contributions to the scattering cross section, so that
     \begin{equation}
          \sigma={4\pi\over{k^{2}}}\sin^{2}\delta_{0}.
     \end{equation}

The effective potential about a point charge in the CBG was calculated
by Hore and Frankel 
\cite{horfra}. The static dielectric function of the CBG is: 
     \begin{equation}
     \label{dil}
          \epsilon(\vec{q},0)=1+\sum_{\vec{p}} 
{4\pi (e^*)^{2}\over{q^{2}\Omega}} 
            \left({F_{0}(\vec{p})-F_{0}(\vec{p}-\vec{q})
\over{-(1/m_b)\vec{p}\cdot\vec{q}+q^{2}/2m_b}}\right),
     \end{equation}
in which  $e^*=2e$ the boson charge, and
 $F_{0}(\vec{p})=(e^{(p^{2}/2m_b-\mu)/k_{B}T}-1)^{-1}$, the
Bose distribution function. It has been shown \cite {alebee2} that 
Eq. \ref{dil} is valid even beyond the simplest random phase appeoximation
assumed in Ref. \cite{horfra}. Eliminating the chemical potential, 
for small $q$ the dielectric function for $T<T_c$ is
     \begin{equation}
          \epsilon(\vec{q}, 0)=1+{4m_b^2\omega_{p}^2\over{q^4}}
             \left[1-\left({T\over{T_c}}\right)^{3/2}\right]
             +O\left({1\over{q^3}}\right),
     \end{equation}
and for $T\rightarrow\infty$ is
     \begin{equation}
          \epsilon(\vec{q}, 0)=1+{1\over{q^2}}
             {m_b\omega_{p}^2\over{k_BT}}
             \left[1+{\zeta({3\over{2}})\over{2^{3/2}}}
             \left({T_c\over{T}}\right)^{3/2}+...\right]+O(q^0),
     \end{equation}
with $\omega_p^2=4\pi (e^*)^2\rho/m_b$,
 and $\rho$ the boson density. 
If the unscreened scattering potential is the Coulomb potential
$V(r)=V_0/r$, then performing the inverse Fourier transforms, one finds that
 for $T<T_c$ \cite{horfra} 
     \begin{equation}
          \lim_{r\rightarrow\infty}V(r)={V_0\over{r}}\exp[-K_sr]
          \cos[K_sr]
          \equiv V_s(r)
     \end{equation}
with 
     \begin{equation}
          \label{scrzero}
          K_s= \left( m_b^2\omega_{p}^2
          \left[1-\left({T\over{T_c}}\right)^{3/2}\right]
          \right) ^{1/4},
     \end{equation}
and for $T\rightarrow\infty$, 
     \begin{equation}
          \lim_{r\rightarrow\infty}V(r)={V_0\over{r}}\exp[-K_nr]
          \equiv V_n(r)
     \end{equation}
with 
     \begin{equation}
          \label{scrinfty}
          K_n=\left({m_b\omega_{p}^2\over{k_BT}}\right)^{1/2}
             \left[1+{\zeta({3\over{2}})\over{2^{3/2}}}
             \left({T_c\over{T}}\right)^{3/2}+...\right]^{1/2}.
     \end{equation}
The $T<T_c$ result is exact for all $r$ at $T=0$.

There are two further important analytical results; the first 
(Levinson's Theorem \cite{calogero}) states that for `regular' potentials 
(which include all
those which we shall be concerned with), the zero-energy phase shift is
equal to $\pi$ multiplied by the number of bound states of the potential.
The second is
the well-known Wigner resonance 
scattering formula \cite{lanlif},
states that for slow particle scattering of a particle
with energy $E$ off a potential with a shallow bound state of binding 
energy $\epsilon\lesssim E$ the total scattering cross section is
     \begin{equation}
          \label{wigner}
          \sigma={2\pi\over{m}} {1\over{E+|\epsilon|}}.
     \end{equation}
We have used this to check that our calculation method works correctly by
comparing our results with Eq. \ref{wigner} for
 various potentials with shallow bound states (Fig. 1).

The zero-energy scattering cross-sections for the potentials 
$V_{n}(r)=-(V_0/r)e^{-Kr}$ and
$V_{s}(r)=-(V_0/r)e^{-Kr}\cos(Kr)$ are shown in Fig. 2a.
 These graphs are plotted
for $V_0=1$; in each case, the equivalent graph for arbitary $V_0$ may
be found by rescaling $\sigma$ and $K$. According to the Wigner formula
(Eq. \ref{wigner}), as $K$ is decreased, when a new bound state appears there
should be a peak in the cross-section, as there will then be a minimum
in the binding energy of the shallowest bound state.
 This is the origin of the
peaks in Fig. 2a, which may be checked using Levinson's Theorem.
 It can also be seen that as $K$ is decreased, the first few 
bound states appear at higher $K$ in the ordinary Yukawa potential; this agrees 
with the intuitive conclusion that the bound states should in general be deeper
in the non-oscillatory potential. Another intuitive expectation which is also
bourne out is that for a given $V_0$ and $K$, the non-oscillatory potential
should be the stronger scatterer; in Fig. 2b it may be seen that this is the
case when $K$ is large enough for neither potential to have bound states (the
difference in cross sections is then in fact about three orders of magnitude.)

Now we address the possible application of our results to the ARPES
linewidth in the cuprates. According to \cite{aleden}, the ARPES peak is
related to photoexctited holes with small group velocity near the
top of the oxygen band. The quantities necessary in order to calculate the scattering cross 
section of impurities (the dopants) in
the cuprates are the static dielectric constant, the effective
mass of the 
bipolaronic carriers, the charge on the scattering centres and the
bipolaron density. The situation is however 
complicated by the anisotropy of the 
effective mass tensor.
The value of the effective mass of the bipolarons 
in the cuprates is 
readily found from the penetration depth \cite{alekab}. 
In BSCCO the in-plane bipolaron mass $m_b$ is about 
$5-6m_e$. The dopants in 
Bi$_2$Sr$_2$CaCu$_2$O$_8+\delta$ are O$^{2-}$ ions, and 
thus the Coulomb potential between a scattering centre and a hole is 
$V(r)=-2e^2/(\epsilon_0r)$. The issue of the dielectric constant is more
contentious; measurements suggest that it may be as high as $1000$
 \cite{epsilon}. The variable
phase method has only been derived for the isotropic problem, so we cannot
apply our theory to make a quantitative conclusion about
 the quasiparticle lifetime
at different temperatures in the cuprates. However, we can provide an
important general conclusion about the relative value of the cross-sections
in the normal and superconducting states.

At zero temperature, the screening wavevector is 
$K_0=(m_b\omega_p)^{1/2}$,
and at a temperature $\alpha T_c$ well above the transition, it is
$K_{\alpha T_c}=(m_b\omega_p^2/k_B\alpha T_c)^{1/2}$. 
Substituting $\omega_p$ and 
$k_BT_c=3.3n^{2/3}/m_b$, we obtain:
     \begin{equation}
          {K_{\alpha T_c}\over{K_0}} = 
          \left( {2.1em_b^{1/2}\over{\epsilon_0^{1/2}\rho^{1/6}\alpha}} 
          \right)^{1/2}.
     \end{equation}
From this, we see that the ratio is only marginally dependent on
 the boson density,
so substituting for $\rho=10^{21}cm^{-3}$, $e$, and $m_e$, we obtain
     \begin{equation}
          {K_{\alpha T_c}\over{K_0}} = 3.0\left((m_b/m_e)^{1/2}
             \over{\epsilon_0^{1/2}\alpha}\right)^{1/2}.
     \end{equation} 
With realistic boson masses 
and dielectic constants, $K_{\alpha T_c}$ and $K_0$, while different, are of
the
same order of magnitude. In the isotropic model, if the screening wavevectors
are such that neither the normal state or condensate impurity potentials have
bound states, with these parameters it would then follow that the quasiparticle
lifetime is much greater in the superconducting state, Fig. 2b.
 We propose that
this effect also occurs in the realistic non-isotropic model, and could then
explain the appearence of a sharp ARPES peak in the superconducting state of
BSCCO. With doping, the screening radius decreases both in the
normal, Eq. 16, and superconducting states, Eq. 14. This explains another 
fascinating experimental observation, namely the strange doping dependence
of the ARPES linewidth. Optimallly and overdoped cuprates, due to the higher 
carrier density, have shorter range scattering potentials with smaller
cross-sections compared with the underdoped cuprates.

In summary, we have calculated the scattering cross section of a Coulomb 
scattering centre in the charged Bose gas both above and below the condensation
temperature. In contrast to the
BCS superconductor, the scattering potential
 in the CBG is different in the normal and 
superconducting states. This is because the coherence length
in the CBG is the same (at $T=0$) as the screening radius \cite{alevor}, 
while in the BCS
superconductor it is a few orders of magnitude larger.
We find that for the realistic parameters, the 
scattering cross section above $T_c$ in the bismuth cuprates might
be around three orders of magnitude larger than at $T=0$. 
We propose that the appearence of a sharp peak in the ARPES spectra of BSCCO
below the superconducting transition and its doping dependence is due to the
condensate screening of the scattering potential. 
We acknowledge valuable discussions with M. Portnoi. CJD was
supported financially in this work by the UK EPSRC.

\centerline{{\bf Figure Captures}}

Fig. 1. Plot of scattering cross section $\sigma_n$ against 
scattered particle momentum $k$ for small $k$ (solid line) and
fit using the Wigner formula (broken line) for potential
$V_n(r)=(-1/r)\exp(-0.55r)$. 
There is good agreement between the numerical
and Wigner results.

Fig. 2. (a) Plots of zero-energy scattering cross sections 
(i) $\sigma_{n}$
and (ii) $\sigma_{s}$ against screening
wavevector $K$ for the potentials (i) $V_n(r)=-(1/r)e^{-K_nr}$ and (ii) 
$V_s(r)=-(1/r)e^{-K_sr}\cos(K_sr)$. (b) Plot of $\sigma_{n}/\sigma_{s}$ for
a range of $K_n=K_s$ in which neither potential has any bound states.
In each case the units are those 
used to derive the phase equation.

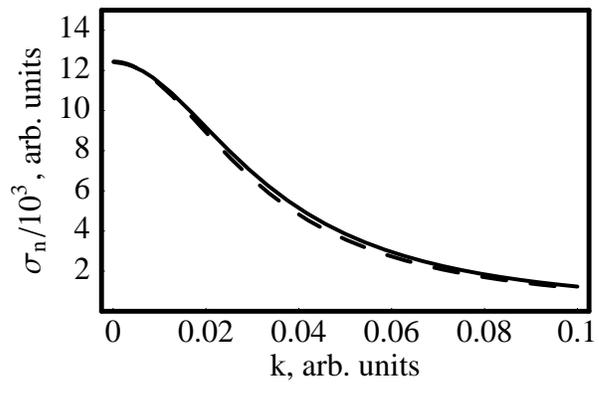
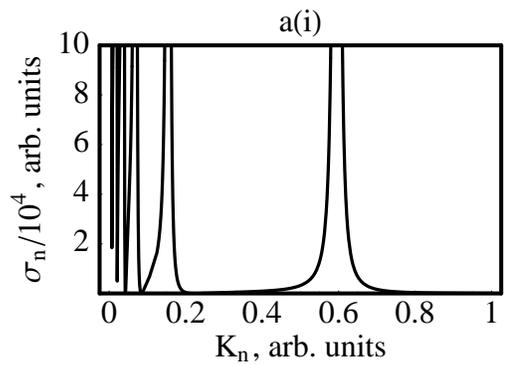
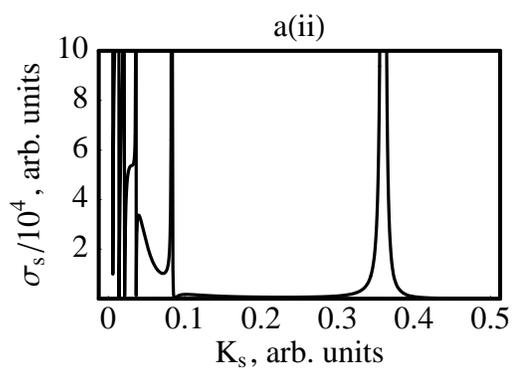
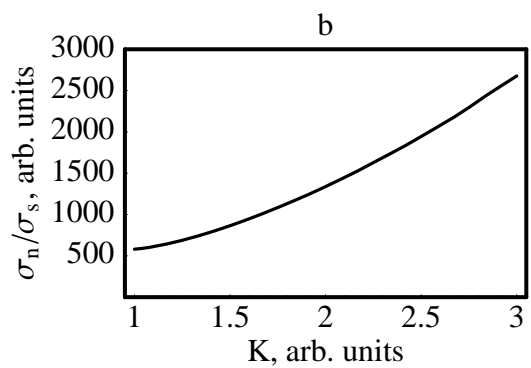
\end{document}